# AN ADAPTIVE MULTIMEDIA-ORIENTED HANDOFF SCHEME FOR IEEE 802.11 WLANS

Ahmed Riadh Rebai[1] and Saïd Hanafi[2]

[1]Electrical & Computer Engineering Program, Texas A&M University, Doha, Qatar
`riadh.rebai@qatar.tamu.edu`
[2]LAMIH Laboratory, University of Valenciennes & Hainaut-Cambrésis, Lille, France
`said.hanafi@univ-valenciennes.fr`

## ABSTRACT

*Previous studies have shown that the actual handoff schemes employed in the IEEE 802.11 Wireless LANs (WLANs) do not meet the strict delay constraints placed by many multimedia applications like Voice over IP. Both the active and the passive supported scan modes in the standard handoff procedure have important delay that affects the Quality of Service (QoS) required by the real-time communications over 802.11 networks. In addition, the problem is further compounded by the fact that limited coverage areas of Access Points (APs) occupied in 802.11 infrastructure WLANs create frequent handoffs. We propose a new optimized and fast handoff scheme that decrease both handoff latency and occurrence by performing a seamless prevent scan process and an effective next-AP selection. Through simulations and performance evaluation, we show the effectiveness of the new adaptive handoff that reduces the process latency and adds new context-based parameters. The Results illustrate a QoS delay-respect required by applications and an optimized AP-choice that eliminates handoff events that are not beneficial.*

## 1. INTRODUCTION

Recent years have been distinguished by a phenomenal growth in the deployment of the IEEE 802.11 [1] Wireless LANs (WLANs) in various environments like universities [2, 3], companies, shopping centers [4] and hotels. This widespread acceptance can be attributed to decreasing infrastructure costs and potential bandwidth that can be offered to the end user. Many believe that the IEEE 802.11 networks are expected to be part of the integrated fourth generation (4G) networks. However, the limited range of Access Points (APs) causes challenging problems. The Mobile Stations (MSs) are required to find and associate with another AP with acceptable signal quality whenever they go beyond the coverage area of the currently associated AP. The overall process of changing association from one AP to another is called as handoff process and the latency involved in the process is termed as handoff latency.

To meet the lofty goal of becoming the next generation networks, the Quality of Service (QoS) for multimedia applications during handoff should be enhanced. The process must be fast enough to ensure continuous connectivity that may be otherwise prevented by several latency sources incurred at different phases of the handoff process. In 802.11 networks, the handoff process can be divided into three phases: *probing* (scanning), *re-authentication* and *re-association*. According to [5, 7] the handoff procedure in IEEE 802.11 normally takes hundreds of milliseconds, and almost 90% of the handoff delay is due to the search of new APs, the so-called *probe delay*. This rather high handoff latency results in play-out gaps and poor quality of service for time-bounded multimedia applications. On the other hand, The MS association with a specific AP is based only on the Received Signal Strength Indicator (RSSI) measurement of all neighbor APs. The MS will disassociate from the AP when the RSSI falls below a predefined threshold. This procedure is based on the conviction that high RSSI is the best indicator of the quality-of-service provided by the selected AP. This naïve procedure, leads to the undesirable result that many MSs are connected to a few APs, while other neighbor APs remain under utilized or idle. The overloaded APs (with high RSSI) will suffer from performance degradation. This raises the need for a better algorithm that takes into consideration the load on the AP and other context-based parameters, as well as RSSI, as part of MS-AP association.

In this paper, we propose firstly a novel Medium Access Control (MAC) Layer handoff mechanism for IEEE 802.11 networks called Prevent Scan Handoff Procedure (PSHP) that reduces the *probe phase* and adapts the process latency to support most of multimedia applications. The PSHP method decreases the delay incurred during the discovery phase significantly by inserting a new Pre-Scan phase before a poor link quality is observed between the MS and its AP. Based on RSSI measurements, the scanned APs in the pre-scan phase will be sorted in a dynamic list. This new phase will be followed by a "prevent handoff" with a new AP offering better conditions than current AP. As a second proposition, we integrate a new and effective AP-selection layer-2 scheme during the handoff procedure based on Neighbor Graph (NG) manipulation. The proposed technique chooses the next most adapted AP from actual Neighbor APs. This choice is performed by means of a new heuristic function that employs multiple-criteria to derive the search. The new network-configuration method differs from the RSSI constrained process by introducing three new network parameters to optimize the next-AP selection during the 802.11 handoff scan phase. The rest of the paper is organized as follows. Section 2 presents an overview of handoff procedure performed in IEEE 802.11 WLANs and related troubles. Related work found in literature is given and discussed in section 3. In section 4, a detailed explanation of proposed schemes is provided. An experimental analysis of our prototype simulation is shown in section 5 followed by the conclusion in section 6.

## 2. THE HANDOFF PROCESS IN IEEE 802.11 WLANS

One of the two modes of operation defined in the IEEE 802.11 standard is the infrastructure-based mode. On these widely used infrastructure-based networks, each MS communicates via a special node called AP. If a MS wishes to send or receive data, it first needs to associate with an AP. The AP acts as a bridge and forwards data packets to appropriate destination. Similarly, all the data packets targeted to MSs are passed through their respective APs. Typically, the AP operates on a specific channel and all the MSs need to compete for the channel using one of the access methods described next. Therefore, for an AP with a single transceiver, only one MS will be able to communicate successfully at any specific point of time. The coverage area of an AP is termed as basic service set (BSS). Extended service set (ESS) is an interconnection of BSSs and wired LANs. The logical medium that interconnects BSSs and wired backbone is termed as distributed system (DS) in the literature. It should be noted that wired interface and APs interface to the DS medium are the part of DS. Infrastructure-based WLAN with three BSSs connected with each other by the DS is shown in Figure 1.

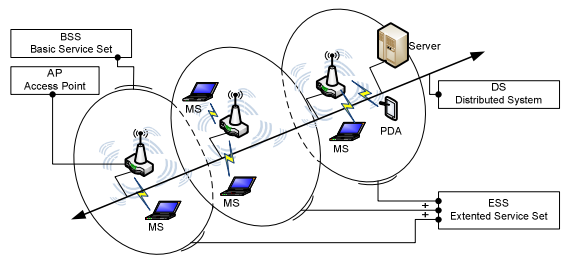

Figure 1. The IEEE 802.11 infrastructure mode

The inter-cell commutation can be divided into three different phases: *detection*, *probing* (scanning) and *effective handoff* (including *authentication* and *re-association*). In order to make a handoff, the MS must first decide when to handoff. However, the IEEE 802.11 standard does not specify any distinct technique to determine when to handoff. The common mechanism is to initiate handoff whenever the Received Signal Strength (RSS) from current AP drops below a pre-specified threshold (termed as handoff threshold in the literature) [5, 8]. Using only current AP's RSS to initiate handoff might force the MS to hold on to the AP with low signal strength while there are better APs in its vicinity. Increasing the handoff threshold does not solve the problem as a larger value drives the MS into performing frequent handoffs. Once the MS decides to make a handoff, the next logical step is to discover the best neighboring AP and re-associate with it. A management frame called *De-authentication* packet is sent, either by the mobile station before changing the actual channel of communication which allows the access point to update its MS-affiliation table, either by the AP which requests the MS to leave the cell. In general, this frame is generated by the mobile station since it detects more quickly the deterioration of the channel quality. After closing the connectivity to the current AP, the MS needs to find potential APs with which to associate. This is accomplished by means of a Medium Access Control layer function called scanning.

There are two types of scanning in the IEEE 802.11 standard: *passive* and *active*. As shown in Figure 2, in the passive scan mode the MS listens to the wireless medium for beacon frames. Beacon frames provide the MS with timing and advertising information. Current APs have a default beacon interval of 100ms [9]. Using information obtained from beacon frames, the MS selects an AP to associate with. During passive scanning, the MS listens to each channel of the physical medium one by one, in an attempt to locate potential APs using the probed channel. Therefore, the passive scan mode incurs significant delay. More technically, the MS commutes from a channel to another one at a regular interval space depending on the setting of *ChannelTime* value. It is indispensable to wait on each channel stated in the *ChannelList* parameter for a time period longer than the *inter-beacon* delays of APs. After scanning all available channels, the MS performs a *Probe* phase (used in active mode) only for the selected AP. As mentioned the polled AP is elected only based on RSSI parameter. The 802.11k group [10] works on improving the choice of the next AP taking into account the network.

In the active scan mode (Figure 3), the MS sends a *Probe Request* packet on each probed channel and waits *MinChannelTime* for a *Probe Response* packet from each reachable AP. If at least one packet is received, the MS extends the sensing interval to *MaxChannelTime* in order to obtain more responses. Contrary, if during *MinChannelTime* the MS does not detect any activity on the channel, the channel is declared *inactive* and the MS passes to the next channel scanning. Thus, the waiting time on each channel is irregular and controlled by two *timers* (not prearranged like the passive scan procedure). When all channels have been scanned, the mobile station collects the information from all available APs and selects the most adequate one to initiate with it the next handoff phase.

The selected AP exchanges IEEE 802.11 authentication messages with the MS. During this phase one of the two authentication methods can be achieved: *Open System Authentication* or *Shared Key Authentication*. The first technique is simply performed by exchanging two authentication packets (request and response) using an access control mechanism based on the MS's MAC physical addresses. It has been shown in [11] the limits of this authentication method by indicating that the access control which is based on mobile's MAC addresses can be easily attacked with software tools that can reconfigure the MAC address of wireless interfaces. The second method assumes the existence of a secret key shared between the station and the access point represented by a *Wired Equivalent Privacy* (WEP) key used also for encrypting data frames. An extra two packets (*challenge - response*) are exchanged during the authentication phase, in which the mobile station must decrypt a text provided by the access point. The method of Shared-Key Authentication requires therefore the exchange of four messages. After that the MS is authenticated by the AP, it sends *Re-association Request* message to the new AP. At this phase, the old and new APs exchange messages defined in Inter Access Point Protocol (IAPP) [12]. Furthermore, once the MS is authenticated, the association process is triggered. The Cell's information is exchanged: the ESSID and supported transmission rates. Only after the association process, the MS will be successfully affiliated with the new AP and can transmit and receive data frames. The initial association starts by exchanging an *Association Request* frame sent by the mobile station that needs to associate. The corresponding AP replays to this request by sending an *Association Response* frame which states whether the association has been accepted or not.

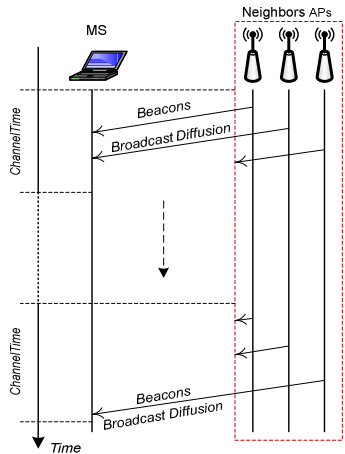
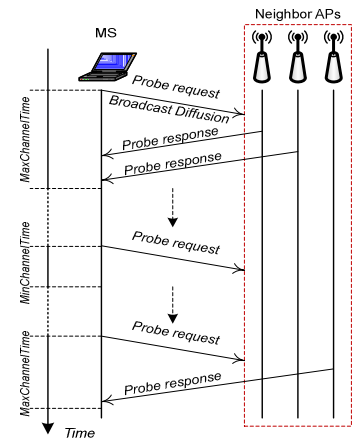

Figure 2. The 802.11 passive scan mode           Figure 3. The 802.11 active scan mode

The total delay incurred during these exchanges is referred as the Layer 2 handoff delay, which consists of probe delay, authentication delay and re-association delay. During these various steps, the MS will not able to exchange data with its AP. Based on values defined by the IEEE 802.11 standard, a station will remain inaccessible by any other entity of the network, for nearly 300 to 500 ms [5, 7]. The scanning phase is considered as the most significant contributor to the overall *handoff latency*. An additional process is involved when the MS needs to change its IP connectivity [14]. In such a scenario, the MS needs to find a new access router. Also, the address binding information has to be updated at the home agent and corresponding agent [15]. In our research work we propose an efficient scheme to decrease the latency involved in finding new neighboring AP, which contributes up to 90% of MAC layer handoff delay.

## 3. RELATED WORK AND RESEARCH OVERVIEW

Numerous schemes have been proposed to reduce the handoff delay in the 802.11 WLANs. In the following, we review the most relevant and representative methods found in the literature. Firstly, we begin with a time investigation for the handoff process to better understand actual difficulties and where to look into. As we have been shown before, the second phase of the handover (*scan* phase) is the most costly in terms of time and traffic. As discussed it is divided into two phases: the *probe* sub-phase and the *channel switching* sub-phase. During these two sub-phases each possible channel must be scrutinized and examined. The latency of the probe sub-phase depends on the adopted scan mode (i.e. passive or active). By assuming the use of a passive scan mode, the average latency of the probe phase depends on the time interval between beacons transmitted periodically by APs and the number of available channels. Explicitly, if the interval between beacons is 100ms of IEEE 802.11b with 11 channels and 802.11a with 32 channels, the average latency will be respectively 1100ms and 3200ms. The switching time incurred while the MS is altering from one channel to another, as it was identified in [16], is negligible and varies between 40 and 150μs. On the other hand, the time incurred with an active scan can be determined by the *MinChannelTime* and *MaxChannelTime* values. Therefore, this quantity can be expressed as shown in Equation 1.

$$N \times MinChannelTime \leq T_{probe} \leq N \times MaxChannelTime \quad (1)$$

where, *N* is the number of available channels. The *MinChannelTime* value should be large enough to not miss the *proberesponse* frames and obeys the formula given in Equation 2.

$$MinChannelTime \geq DIFS + (CW \times SlotTime) \quad (2)$$

where, *DIFS* is the minimum waiting time necessary for a frame to access to the channel. The backoff interval is represented by the contention window (*CW*) multiply by *SlotTime*. In other words, the parameter *MinChannelTime* represents the maximum time for sending a frame. Once this time is elapsed, the MS should receive a response from the access point, and so, will increase the waiting time to *MaxChannelTime* for other potential responses. Otherwise, the station considers that there is no AP on the scanned channel, or other traffic are competing the channel access with the expected management frame.

Regarding the third phase of the handover procedure that allows the MS's identity verification, the delay is changeable. According to the security used, the authentication process can be more or less long. In an untrusted system, only two Authentication frames are exchanged, with their respective 802.11 acknowledgments. Using a secure system, such as WEP, four frames must be exchanged. The latency of the authentication phase is proportional to the number of messages exchanged between the AP and the MS. For example, public systems recently deployed WLAN (e.g. nespot in Korea [17]) use the authentication scheme of the IEEE 802.11x. Therefore, the authentication phase is expected to become an issue much more interesting in future 802.11versions.

The phase of association or re-association process that comes to the end for the 802.11 handover takes place through the exchange of two frames (Association Request and Association Response), both messages will be acquitted. The duration of this phase, such as the authentication phase, is limited to the medium access time which depends on the traffic in the cell (such management frames have no special priority) and to the delay of their transmissions. In [9] the delay of these last two phases was estimated to less than 4ms in absence of a heavy traffic in the new selected cell. The total handover latency is expressed in Equation 3.

$$T_{Handover} = N \times (T_{switch} + T_{probe}) + T_{authentication} + T_{association} \quad (3)$$

where, *N* is the number of available channels depending on the *ChannelList* parameter. In practice, and based on Equation 3, a handover performed on the standard 802.11 network can theoretically have values ranging from 114 ms to 940 ms (for $N = 11$). This value is very high and not acceptable for most of applications with QoS requirement (e.g. voice frames that are time-bounded must be received every 50 ms).

In [18] authors propose an innovative solution to optimize the AP's exploration during the scan phase based on the use of sensors operating on the 802.11 network. These sensors are arranged in cells and spaced 50 to 150 meters.

These sensors have a role to listen to the network using beacons sent periodically by in-range APs. Each sensor is able to identify the nearby access points that are available. When the MS should change its actual cell, it performs a pre-scan operation which involves the sending of a request query to the sensors. Only sensors that have received this request (in range of the MS) react by sending the list of APs that they have identified. Each sensor responds by using a contention window calculated proportionally to the signal strength of the received request message. We figure out that this solution is effective in terms of the next-AP choice and the consequent results have improved significantly the standard handoff scheme. However, it is very expensive and has an extra cost by causing an additional load of unnecessary network traffic due to the sensor use. Moreover, this method is a non compliant solution with the actual 802.11 networks and requires radical changes to adapt it.

In [19], a new handoff scheme called *SyncScan*, is proposed to reduce the probe delay. Unlike the existing probe procedures defined in IEEE 802.11, *SyncScan* allows a MS to monitor the proximity of the nearby APs continuously. In other words, the MS regularly switches to each channel and records the signal strengths of the channels. By doing so, the MS can keep track of information on all neighbor APs. Essentially, this technique replaces the existing large temporal additional costs during the scan phase by a continuous process that passively monitors the presence of access points in other channels. The absence delay of the MS with its current channel is minimized by synchronizing short listening periods of other channels with regular periodic beacon transmissions from each AP. Moreover, through continuous monitoring the signaling quality of multiple APs, a better handoff decision can be made and the authentication / re-association delay can be also reduced.

The authors synchronize the MS with the transmission of beacons from the APs on each channel. By switching regularly and orderly on each channel, the MS reduces its disconnection delay with its actual AP. However, the *SyncScan* process admits a hidden cost. While it removes the scan phase delay, it adds regular additional interruptions between the MS and its actual AP. Specifically, when the MS examines other channels it cannot send or listen to its own AP. As results, the MS may miss packets that were sent when exploring other channels. These errors are very costly in terms of frame loss and performed retransmissions especially for time-bounded applications. Moreover, this extra charge will always affect all MSs even those that will never proceed to a handoff.

In [20], the authors proposed a *selective scan* technique in the IEEE 802.11 WLAN contexts that support the IAPP protocol [12] to decrease the handover latency. This mechanism reduces the scan time of a new AP by combining an enhanced Neighbor Graph (NG) [21] scheme and an enhanced IAPP scheme. If a MS knows exactly its adjacent APs, it can use selective scanning by unicast to avoid scanning all channels. During handoff, a MS does not scan all channels. Instead, it selectively scans few potential APs with unicast based on the NG provided by a NG Server called RADIUS server [22]. They enhanced the NG approach by putting the MS to power-saving mode (PSM) to pre-scan neighboring APs. Then they further derived selective scanning with unicast in power-save mode, pre-registration of IAPP, and frame forwarding-and-buffering mechanisms. Selective scanning allows a MS to only try potential handoff targets. Pre-registration allows early transfer of the MS security context from its old AP to new AP. The forwarding-and-buffering mechanism is to solve the packet loss problem during the handoff process. This solution reduces, in a remarkable way, the total latency of the handoff mechanism. On the other hand, it requires that the MS must have knowledge of the network architecture, it must know exactly the APs which are adjacent for it to be able to employ selective scan and to avoid scanning all channels. In addition, we should take into account the number of packets added by the IAPP that may affect the current traffic. Moreover, we note that all data packets have been sent to the old AP and then routed to the new selected AP before the link-layer is updated, which corresponds to a double transmission of the same data frames in the network. Thus, it greatly increases both the collision and the loss rates in 802.11 wireless networks.

In [23], the authors proposed two changes to the basic algorithm of IEEE 802.11 which reduce significantly the handover average latency using inter-AP communications during the scan phase. In the first proposed scheme, the additional costs incurred during this phase are reduced by forcing the potential APs to send their probe response packets to the old AP and not to the MS which sends the probe request. Therefore, the MS will avoid the waiting delay of *MinChannelTime* or *MaxChannelTime* as performed in the classical IEEE 802.11 approach. Consequently, the MS avoids the packet loss of data without any additional cost in the network. This efficient and fast handoff process is called Fast Handoff by Avoiding Probe wait (FHAP). As explained, the probe wait is avoided by forcing all the neighboring APs operating on the examined channel to send the probe responses to the previous AP using the IAPP protocol. The MS just switches to all the channels and sends the probe request. After the probe phase the MS switches back to its actual AP and it receives the probe responses after sending a request. The discovery phase ends at this time and the MS resumes re-authentication process with the new AP.

Three drawbacks related to the FHAP approach can be noted down. Firstly, the MS should receive packets (probe response messages from potential APs) from its old AP. This implies that the handover threshold must be adjusted so that the MS can communicate with the old AP after the probe phase. Secondly, the problem of non-delivery probe response packets from potential APs to the old AP should be addressed. Finally, as the probe response packets are received via the current AP and not on their respective channels, the MS will not be able to measure the instantaneous values of RSSI and therefore evaluate the quality of the visited channel (which is possible only if the reception is done on the same channel).

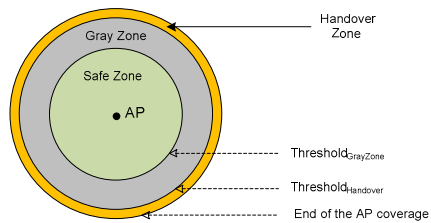

Figure 4. Sub-Zone partitioning in APFH [23]

In [23, 24], the authors have improved their technique FHAP by proposing a new mechanism called Adaptive Preemptive Fast Handoff (APFH). The APFH method requires that the MS predetermines a new AP before the handover begins. Then, the handover threshold is reached, the MS avoids the discovery phase and triggers immediately the re-authentication phase. This process will reduce the total handover latency by decreasing its value to the re-association/authentication delay. Since the authors did not specify how the MS preselects a new AP, we can figure out that the *SyncScan* mechanism [19] presents a solution to this problem. The new adjustments achieved in APFH technique provide a better preemptive scan phase of APs. As shown in Figure 4, the APFH method splits the coverage area of the AP depending on the signal strength in three areas: safe zone, gray zone and handover zone. As its name indicates, the safe zone is the part of the coverage area where the MS is not under a handover threat. Consequently, the MS does not trigger the discovery phase and the data transfer is accomplished normally. The gray area is defined as an area where the handover probability is high. The MS begins collecting information on a new best AP once it enters the unsure zone. The maximum selected speed of MSs for the simulations is 15 m/s as in [25]. In conclusion, since the first proposed scheme FHAP as discussed does not meet the QoS constraints of multimedia applications by receiving all response frames on the old AP, the authors presented a second mechanism called APFH that removes the entire handover latency and respects these strict constraints of VoIP frames.

Many research works were done on the network-layer regarding the challenge to support the mobility in IP networks. New features have been proposed and added to the standard – i .e. IPv6 [14, 15]. However, the best handoff techniques that minimize the scan phase delay are performed at the MAC-layer of the 802.11 standard. By minimizing this phase latency, then the number of lost packets is reduced and the MS will be not reachable only for a limited time.

## 4. PROPOSED SCHEMES

### 4.1. Prevent-Scan Handoff Procedure (PSHP)

First, the typical handoff latency in IEEE 802.11b with IAPP network may take a probe delay of 40 to 300ms with a constant IAPP delay of 40ms [26]. To allow the IAPP protocol to reduce this delay, we impose that the MS must authenticate itself with the first AP of the ESS. However, the IEEE 802.11 standard neither requires that authentication must immediately proceed to association nor that authentication must immediately follow a channel scan cycle. The IAPP based pre-authentication [27] is achieved even before MS enters into the discovery state, thus, it does not contribute to the handoff latency.

As a first modification, we propose to define a new threshold other than the existing handoff threshold in the 802.11 standard. We call the new threshold: Preventive Received Signal Strength Indicator which is termed by ($RSSI_{prev}$) and defined in the given Equation 4.

$$RSSI_{prev} = RSSI_{min} + (RSSI_{max} - RSSI_{min})/2 \qquad (4)$$

According to our implementation and the tests that we carried out, $RSSI_{max}$ indicates the best link quality that can exist between the MS and its AP. As its name implies, the $RSSI_{prev}$ is a value of the *link quality* above which the MS is not under the threat of imminent handoff. In the proposed approach, the algorithm starts to detect the mobility of a MS when the RSSI value of the current AP degrades and reaches the $RSSI_{prev}$ threshold, after which the MS starts to seek a new AP which can offer a better link quality.

As described before the best mechanisms – as the *SyncScan* mechanism [19], the proposed selective scan [20], and the APFH technique [23] – are imposing that the MS must predetermine a new AP before the start of handoff. Thus, when the handoff threshold is reached, the MS jumps the discovery phase and starts directly the re-authentication phase. This procedure reduces the overall handoff latency. Therefore, most of the operations related to handoff are executed before that a handoff is triggered, including the selection of the next AP and the transfer of MS's context. For each *SyncScan* procedure, the MS must switch to a specific channel until it receives the corresponding beacon, then it switches back to the original channel. So, for each channel the *SyncScan* latency is given by Equation 5 as follows:

$$SyncScan_{delay} = 2 \times T_{switch} + T_{wait} \qquad (5)$$

where, $T_{switch}$ is the switching delay from one channel to another and $T_{wait}$ is the time required to recover the beacons issued by the APs running on a given channel. The total delay of the handoff scan depends on the number of channels to be scanned.

*a) Association procedure*

We present a novel approach that provides an enhanced technique for the *preemptive* scan of APs. Indeed proposed technique requires carrying out a scan (or pre-scan) even before triggering a handoff. We continuously maintain the information concerning at most the best few nearby APs in a dynamic list sorted according to the descending order of the best RSSI values. This list, which is updated after each pre-scan, reduces significantly the scanning delay to nearly zero. Each MS maintains its classified AP list. Using this list, the MS does no longer need to carry out a full scan when a handoff is initiated. Rather, it directly selects the AP in the first position of the AP list and performs an association request. Note that an association request will be accepted only if the RSSI of the first AP, in the dynamic list, has a value greater than both the handoff threshold and the actual RSSI measured with the current AP. In other words, this request is accepted only if the first AP of the dynamic list offers to the MS a better link quality better than offered by its current AP and also sufficient to continue operation without losing connectivity with the other entities of the network. If the association with the first AP fails, then the dynamic list is purged and the MS carries out a new pre-scan cycle.

*b) A new Pre-scan process*

During a pre-scan process, the MS must switch channels and wait for beacons from potential APs, which produces additional temporal costs composed by switching time between channels and waiting time on each one. Consequently, for each channel we calculate the total time of pre-scan using the following Equation 6.

$$T_{pre-scan} = N \times (T_{switch} + T_{wait}) \tag{6}$$

Where, $N$ is the number of available channels, $T_{switch}$ is the switching delay from one channel to another and $T_{wait}$ is the time required to receive the potential beacons on a given channel. Despite $T_{switch}$ and $T_{wait}$ have relatively small values; they are still greater than the maximum retransmission time of 802.11 frames (4ms). Therefore, time-bounded packets may be dropped since the MS is unable to acknowledge them. To overcome this drawback, we modified the MS build-in algorithm to announce entering a Power Saving Mode just before switching channels [28]. This causes the AP to buffer packets until the MS returns to its channel and resets the PSM mode. Since these buffers will not be overfilled during the PSM mode (very short in duration), they are quickly emptied when the MS finishes the pre-can process and returns to normal mode.

The *pre-scan* is programmed so that it does not disturb the existing traffic flow between the MS and its AP. After each execution of the pre-scan, the MS must check its current RSSI value. Once the MS associates with a new AP, then it initiates a pre-scan again. The flowchart in Figure 5 illustrates the new pre-scan procedure of the proposed handoff scheme.

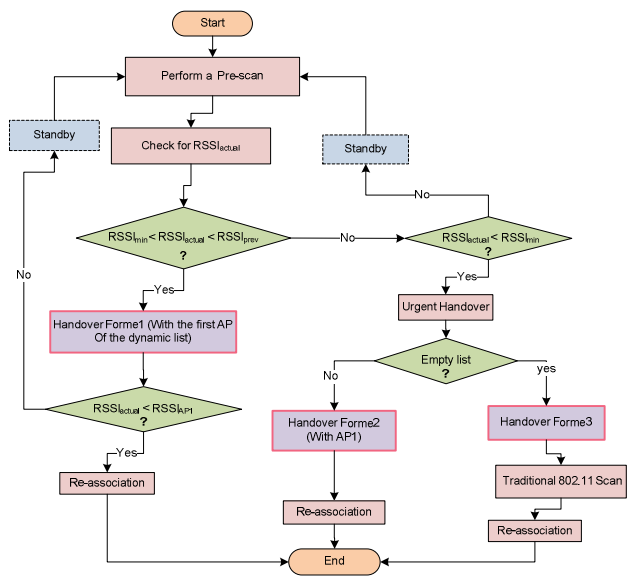

Figure 5. The new PSHP process Flowchart

*c) Operation of the new PSHP mecanism*

Figure 6 presents the new state machine for a MS showing the various amendments that we have added to the basic algorithm. Throughout its activity/mobility, the MS can be in one of several states and has various RSSI values. The variation of the RSSI value can be also due to other factors, for example the channel conditions, interference, and AP traffic loads. In the following, we explain our new approach by detailing the various states that a MS may be faced, and conditions that trigger the transition from one state to another. When a MS is initiated in the network, it firstly associates with an active AP. The MS is required to directly proceed to an authentication (called Pre-Authentication) with all other APs in the same ESS. Following the pre-authentication phase and when needed (scan phase), the MS will notify its current AP that it is entering the power-saving mode so that the AP can buffer the incoming data for the corresponding MS. The MS carries out a periodic active scan (each α ms), called a pre-scan phase. We decide that this cyclic pre-scan will not depend on the existing traffic category between the MS and its current AP since it requires a deep cross-layer knowledge of the traffic type. In other words, the MS performs the planned pre-scan mode when lower or higher priority traffic is transmitted on the channel. Otherwise, the proposed mobility technique will have a hidden and costly delay because it can not start until a deep packet classification based on the application data inside IP packets is performed. In fact, carrying out such classification before a pre-scan does not affect the QoS constraints since TCP will retransmit missing packets. As well, the effect may be worse on RTP/UPD traffic. If we choose to trigger a pre-scan mode only when low priority traffic is adopted between the MS and its actual AP and there were QoS packets just transmitted on the network, the pre-scan phase will never happen and our implementation will be not valid and totally worse. Thus, we simplify the algorithm by making the MS enters the PSM mode whenever RSSI crosses the threshold and a planned pre-scan is launched.

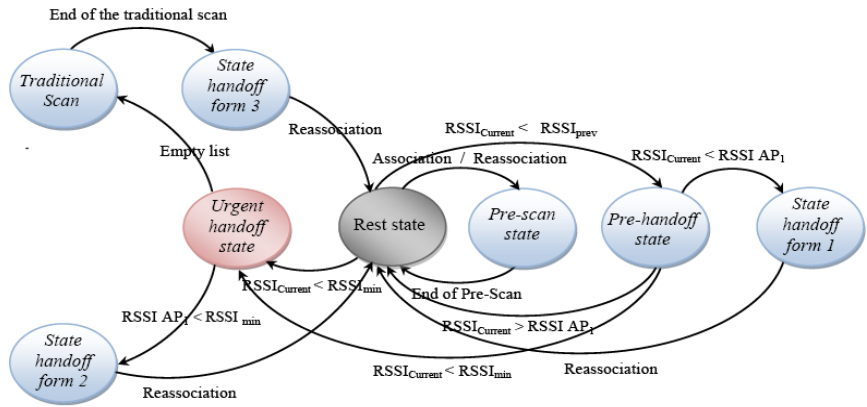

Figure 6. State machine of the PSHP procedure performed by the mobile station

The major advantage of the proposed scheme is that a MS will seek periodically for a new AP offering a better quality of link for forthcoming transmissions between the MS and its associated AP. The periodicity of the pre-scan phase is referred by the parameter α which is defined in the following Equation 7:

$$\alpha = [(T_{switch} + MaxChannelTime) \times N] \times 1.5 \qquad (7)$$

Where $T_{switch}$ is the switching delay from one channel to another, *MaxChannelTime* represents the maximum waiting time to collect all potential probe responses from other APs, and *N* is the number of available channels. The *N* value varies depending the standard from 11 channels in 802.11g to 32 channels in 802.11a (e.g. *N*=13 in IEEE802.11b [29]). The periodicity α value is chosen as a manner to ensure that the MS finishes the actual pre-scan cycle and leaves the PSM mode to join back the active mode and receive data packets held by the actual AP before performing the next pre-scan cycle. During the pre-scan period, the proposed algorithm will drive the MS to collect and keep valued information related to each potential AP in the network in a dynamic list. Initially, this list is empty and will be updated after the first pre-scan cycles to include at least one AP with which the MS can be associated if the link quality with its actual AP will attain the handoff threshold. The new technique chooses a maximum of six APs to be saved in the list, since most WLAN infrastructures adopt a hexagonal deployment of AP cells. Thus, the result of the pre-scan cycle is an ordered list of the nearby APs according to their RSSI values. This process is periodically generated to update the dynamic list. Therefore, this repetitive deployment enables the MS to be always reorganized facing to the active network state by keeping a dynamic list linked to the events that occured in the nearest past.

In the proposed handoff scheme we enumerate three forms of handoff that can be happened depending on network conditions. Initially, the MS is in standby state as shown in Figure 6. If the RSSI value associated to the current AP degrades and reaches the $RSSI_{prev}$, then the MS switches to the pre-handoff state to check its dynamic list. It will try to find out a new AP with a corresponding RSSI value higher than the actual one. If such value exists, the MS switches to a 'handoff form1' state and performs a re-association procedure with the chosen AP. Otherwise, the MS returns to its standby state. We notice if such case is achieved ('handoff form1' state) the total latency of the handoff mechanism is reduced to a cost almost equal to that of the re-association delay. If the measured RSSI value with the current AP is deteriorating suddenly and reaches the minimum bound (handover threshold), then the MS passes directly from the standby state to the 'urgent handover' state. In such state the MS must decide whether to perform the second or the third form of handover depending only on the instantaneous data of the dynamic list. If the first AP in the list has a RSSI value greater that the handover threshold, then the MS switches to the 'handoff form2' state. It performs a re-association process with the selected AP and returns to the standby state. If such case does not exist, the MS switches to the 'handoff form3' state in which it carries out a classical 802.11 handoff with a traditional scan procedure. We figure out in 'handoff form2' process, the overall handoff latency is equal also to the re-association delay and the MS chooses an AP from the list which guarantees a minimum channel quality required to transmit data packets. In the third handoff form the MS joins the new AP after executing a standard scan process and returns to the standby state. However, this state is rare and very occasional in practice (the list is rarely empty after carrying out pre-scan cycles).

The main advantage of the proposed technique is its autonomy since it follows instantaneous network variations and takes appropriate decisions accordingly. This allows a faster and more adequate handover occurrence and a channel quality improvement. In addition, the periodic scan presents another opportunity to improve the link quality of the MS with its AP.

### 4.2. New enhanced technique for a best next-AP selection

In this sub-section, we show a novel and effective layer-2 AP-selection add-on technique for the handoff procedure. The proposed method chooses the most adapted AP from actual Neighbor APs for the next handover occurrence. This choice is performed by means of a new heuristic function that employs multiple-criteria to derive the search. We point out that the standard procedure is based only on the RSSI value considered as the best indicator of the quality-of-service provided by the AP. This naïve procedure, leads to the undesirable result that many MSs are connected to a few APs, while other neighbor APs are underutilized or completely idle. The overloaded APs (with high RSSI values) will suffer from performance degradation. This inefficacity raises the need of a better algorithm which takes into consideration the load on the AP, as well as the RSSI indicator, as part of associating a new MS to the AP. The load balancing problem, as part of handoff, has not been adequately addressed in the literature. In [30], the authors argued that the login data with the APs can reflect the actual situation of handovers given discrete WLAN deployment. As an example, two WLANs may be very close to each other but separated by a highway or a river. In such case, the user will never move across to the other WLAN. Conversely, if the user is moving fast (e.g. in a train), handover may need to take place among WLANs that are far apart, i.e. among non-neighbor APs. Thus, the user connection history allows us to better predict the probability of the user's next movement.

*a) New decisional parameters*

We propose a new network-configuration method that differs from the RSSI constrained process [13] by introducing three new network parameters to optimize the next-AP selection during a WLAN handoff procedure. The first parameter is called $MS_i$ and indicates the number of MSs associated with the potential neighbor $AP_i$. Thus, the new parameter exploits the overload of $AP_i$ as a handoff indicator. A handover occurrence with an overloaded AP may not be beneficial for both the MS and the chosen AP. The second parameter is called $CNX_i$ which is a history-based factor that counts handoff occurrences between the actual AP and the potential next $AP_i$. This counter is incremented by one each time a handoff occurs between AP and $AP_i$. This parameter is adopted to select the neighbor $AP_i$ with the maximum $CNX_i$ value as the best candidate for the next handover against other neighbors. It includes the location and other context-based information useful for the next AP-selection. The third parameter is $EXT_i$ which reproduces the number of APs which are neighbors of the potential $AP_i$ chosen for the next handoff process with the current AP. In other words, $EXT_i$ is the number of 2-hop neighbors – denoted by $AP_k$ – of the current AP through a direct neighbor $AP_i$. This "look-ahead" parameter is added to improve the choice of the potential $AP_i$ to maintain long-term connections and maximize the handoff benefit for new affiliated MSs.

Now, we describe necessary modifications in the MS-AP communication protocol regarding the algorithm implementation. The following new fields must be added to the Beacon and Probe response frames to transmit the additional information needed by the algorithm:

- $MS_i$, the number of mobile stations (MSs) associated with $AP_i$.
- $CNX_i$, the number of actual handoff occurrences between the current AP and desired $AP_i$ neighbor.
- $EXT_i$, the number of 2-step neighbor $AP_k$ of actual AP through its direct neighbor $AP_i$.
- $RSSI_i$, the RSSI value of the incoming Probe Request from neighbor $AP_i$.

As shown in Figure 7, the MS will choose the best $AP_i$ for the actual handoff process after receiving all Probe Responses with the required information.

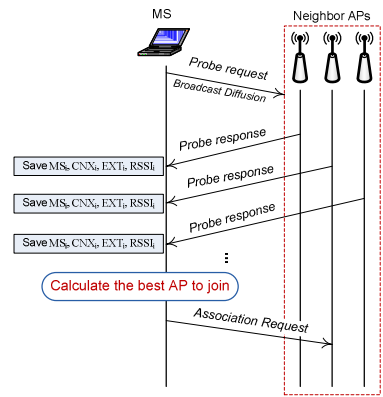

Figure 7. The new AP-selection procedure

*b) Mathematical formulation*

In this paragraph, we introduce a new numerical optimization approach for the next-AP selection in the IEEE 802.11 handoff procedure. We start by introducing the following assertions:
- A time limit factor is needed because of the MS mobility.
- The next-AP selection can be formed as an *assignment problem*: how to allocate MSs with the given set of APs.
- The measured RSSI values can be different for two MSs allocated to the same AP.

We will adopt the following notation:
- $I$ : is the set of APs
- $J$ : is the set of MSs
- $N(AP_i)$ : is the set of the direct neighbors of $AP_i$
- $O_{ik}$: is the number of handoffs that are performed between $AP_i$ and $AP_k$

Then, the matrix $A$ is representing MS-AP affiliations:

$$A_{ij} = \begin{cases} 1 & \text{if the mobile station MS}_j \text{ is associated to the AP}_i \\ 0 & \text{otherwise} \end{cases}$$

We assume that the current mobile station $MS_{j°}$ is assigned to the $AP_{i°}$, and our goal is to assign the current $MS_{j°}$ to the best new $AP_{i*} \in N(AP_{i°})$ such that:

1) Max $\{RSS_i : i \in N(AP_{i°}) \text{ with } RSS_i \leq Threshold\}$
2) Max $\{|N(AP_i)| : AP_i \in N(AP_{i°})\}$
3) Max $\{O_{i°i} : i \in N(AP_{i°})\}$
4) Min $\{\sum_{j \in J} A_{i,j} : i \in N(AP_{i°}) \text{ with } \sum_{j \in J} A_{i,j} < m\}$ (where $m$ is the maximum load of an AP).

Now we define the variables $x_i$ as: $$x_i = \begin{cases} 1 & \text{if the current station } MS_{j°} \text{ is assigned to the } AP_i \\ 0 & \text{otherwise} \end{cases}$$

This problem can be formulated as a multi-objective optimization problem:

$$\begin{aligned} & Max \sum_{i \in I} RSS_i x_i \\ & Max \sum_{i \in I} |N(AP_i)| x_i \\ & Max \sum_{i \in I} O_{i°,i} x_i \\ & Min \sum_{i \in I} \sum_{j \in J} A_{i,j} x_i \end{aligned}$$

*Subject to*
$$\sum_{i \in I} x_i = 1$$
$$\sum_{i \in I} RSS_i x_i \leq Threshold$$
$$\sum_{i \in I} \sum_{j \in J} A_{i,j} x_i < m$$
$$x_i = 0 \quad \text{for } i \notin N(AP_{i°})$$
$$x_i \in \{0, 1\} \quad \text{for } i \in I.$$

## 5. SIMULATION RESULTS AND DISCUSSIONS

### 5.1. The proposed PSHP evaluation

*a) Parameter setting*

In this section, the performance of the proposed scheme PSHP is evaluated and compared to the basic handoff scheme (currently used by most network interface cards) and other significant works founded in [9, 19, and 23]. The handoff latencies of all schemes for different traffic loads are presented. This is followed by discussion on the total amount of time spent on handoff for all schemes. The effect of the proposed schemes on real time traffic is explored and weighed against the basic handoff scheme. We used C++ to simulate the new 802.11 handoff versus other described techniques. The IEEE 802.11b [29] networks are considered for testing the schemes. The total number of the probable channels is assumed to be 11 channels (number of all the legitimate channels used in USA for 802.11b). We employed a total of 100 APs and 500 MSs to carry out the simulations. The other parameters are outlined in Table 1.

Table 1. Simulation Parameters

| Parameter | Value |
|---|---|
| Speed of MS | 0.1 – 15 m/s |
| Mobility Model | Random Way Point |
| MinChannelTime / MaxChannelTime | 7/11 ms |
| Switch Delay | 5 ms |
| Handoff Threshold | -51 dB |
| Pre-Scan Threshold | -45 dB |

In general, all the solutions suggested for optimizing the handoff process aim to reduce the total latency below 50ms [31] mainly for multimedia applications. The proposed PSHP solution aims to be conforming to this restriction by reducing the total handoff delay incurred in 802.11 WLANs. We choose a free propagation model for the mobile stations. Thus, in performed simulations the received signal strength indicator value is based on the distance between a MS and its AP (RSSI-based positioning) as shown in [32]. The relationship between distance separating a MS and its AP and the received signal strength is described in Equation 8.

$$P_r(d) = P_{0\,(transmitting\ power)} - 20\log_{10}\frac{4\pi d}{\lambda}\,[dBm] \quad (8)$$

Where $\lambda = c/f$, $f$ is the used transmission frequency and $c$ is the velocity of light (the propagation speed of waves). The adopted mobility model is based on the model of random mobility "Random Way Point Mobility Model" presented in [33]. The same moving model has been also adopted in other algorithms [19, 20, 21, and 23].

*b) Simulation Results*

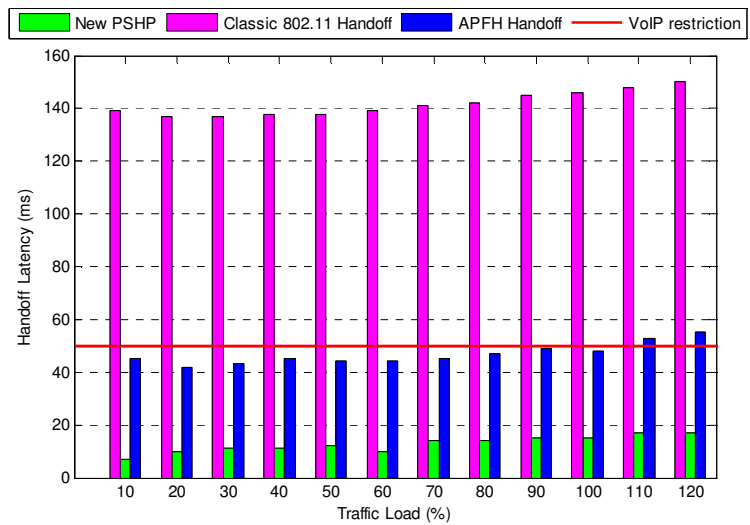

Figure 8. Handoff Latency versus Traffic Loads

Figure 8 shows the average handoff latency against different traffic loads for the three tested schemes. The APFH scheme achieves 67.62% delay improvement while the new PSHP method attains 95.21% improvement versus the basic 802.11 handoff scheme. The handoff latency of the classical approach is consistent with the simulation results in [9] with similar parameters. Also, we point out by observing Figure 8 that the average handoff latencies for the PSHP and APFH schemes are both under 50ms which is well within VoIP constraints. However, PSHP performs the best and the minimal handoff delay compared to the APFH [23]. This remarkable improvement is reached since the new procedure performs a cyclic pre-scan phase before carrying out a handoff and most of handoffs are accomplished early by detecting the premature quality deterioration. As in [19, 23] the traffic load is computed by dividing the number of active MSs (the MSs having data to transmit) over the maximum number of MSs transmitting on one AP's cell. The maximum number of active MSs is equal to 32 in IEEE 802.11 WLANs.

Based on the given results in [5, 19, and 21] of related handoff techniques, we draw the following Table 2 resuming the total handoff delays for corresponding proposed mechanisms. We figure out a significant reduction achieved by the new PSHP algorithm compared to other solutions, and more specifically with the basic handover mechanism. We also note that the solution called *SyncScan* has an important reduction and can also satisfy the time-bounded applications. However, the selective scanning method occasionally exceeds the required QoS limits. This result is due to the inefficacity of the NG graph technique to manage all network topology changes due to the continuous MS mobility. Regarding the new handoff method, and as expected, the total latency is reduced only to the re-authentication phase (≈11ms). This delay can reach more (18ms) because in some simulated cases a handoff occurrence is triggered while a pre-scan cycle did not finish.

Table 2. Average latencies of different handoff procedures

| Scan Technique | Total Latency |
| --- | --- |
| SyncScan [19] | 40±5ms |
| Selective Scan [21] | 48±5ms |
| APFH [23] | 42±7ms |
| Traditional 802.11 handoff | From 112ms up to 366ms |
| Proposed PSHP | 11±7ms |

Figures 9 and 10 evaluate the performance of the APFH technique – the best known solution in literature – and the new PSHP scheme against VoIP traffic. The packet inter-arrival time for VoIP applications is normally equal to 20ms [34], while it is also recommended that the inter-frame delay to be less than 50 ms [31, 34]. This restriction is depicted as a horizontal red line at 50ms in Figures 9 and 10. A node with VoIP inter-arrival time is taken and the corresponding delays are shown. The vertical green dotted lines represent a handoff occurrence. The traffic load for the given simulations was fixed to 50% and the number of packets sent to 600 (≈ 2.5 s).

We remark that handoff occurrences are not simultaneous for the two simulated patterns. The MSs adopting the new PSHP algorithm detect the quality deterioration with their corresponding AP earlier than the APFH process. We note that both techniques respect the time constraint of real-time applications on recorded inter-frame delays without exceeding the required interval (50 ms). However, this constraint is better managed by the new approach and the inter-packet periods are more regular and smaller. As discussed before, the handoff latency for PSHP scheme is just the re-authentication delay if all handovers occur under the first or second form. If the third form of handover is performed, then the latency will be equal to the delay incurred in legacy 802.11 scanning all channels in addition to the re-authentication delay. However, most of PSHP handoff occurrences are carried out using the first and the second form.

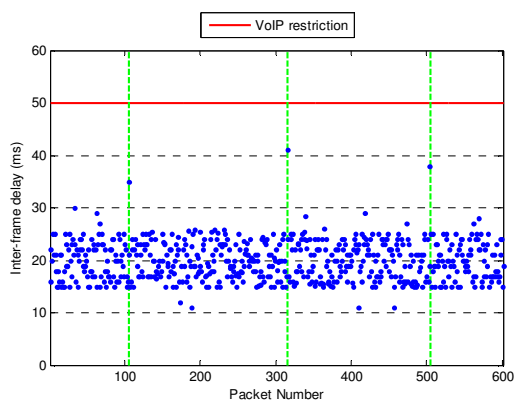 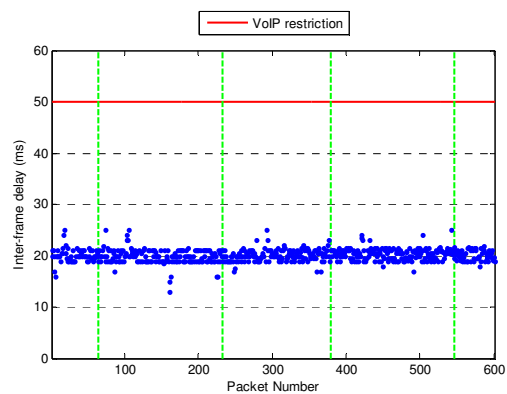

Figure 9. Inter-frame Delay in APFH [23]       Figure 10. Inter-frame Delay in PSHP

To ensure this last assertion we present in Figures 11 and 12, respectively, a count of handoff occurrences for both APFH and PSHP schemes according to the traffic load and the detailed number of the various handoff forms related to the new PSHP technique. We set the simulation time to 10s for each considered traffic load.

By comparing values obtained by the two algorithms in Figure 11, we easily point out that the APFH technique [23] performs less handovers in the network than the proposed PSHP scheme. This result can be explained by the adoption of the new form of preventive Handover (called Form 1). Using this new form, a MS will not wait for a minimum quality recorded equal to the handoff threshold to trigger a handover. This new technique detects early the link quality deterioration with its current AP and performs a switching with a new AP which improves link conditions. Therefore, the periodic pre-scan adopted by the new technique offers new opportunities to enhance the link quality between a MS and its AP and a significant reduction of the total handover delay. Indeed, with the pre-scan cycle the MS can discover other APs that have a better value of RSSI than provided by the current AP and provide the means to make more intelligent choices before and during a handover. The new algorithm PSHP has a better choice for the next AP by collecting periodic RSSI measurement. Thus, the decision is earlier and more beneficial when a handover is performed (rather than relying on a single sample as in usual schemes). Consequently, the extra number of PSHP occurrences versus APFH procedure happenings is compensated by an early choice of next AP with a better offered quality.

In Figure 12, the vertical red lines represent the executed number of Form 1 handoffs. Blue lines represent the number of handoffs taken under the second and third form, i.e. urgent handoffs. Recall that handoff under the first form is started when the RSSI value degrades below the $RSSI_{prev}$ and above the handoff threshold. Handoffs of the second and third form start only if RSSI value is degraded below the handoff threshold. In Figure 12 we figure out for most traffic loads, urgent handoffs occur less frequently than handoffs of Form 1. We also state that the proposed algorithm presents true opportunity to improve link quality since most of handoff occurrences are executed before that the RSSI value degrades below the handoff threshold. Accordingly, we conclude that almost half of accomplished handoffs are done under the new first form, which explains the delay reduction of PSHP since the first form decreases the related latency considerably and improves the link quality between the MS and its current AP.

Table 3 shows the average probability of data packets being dropped and caused mainly by handoff procedure for the three schemes (APFH, PSHP, and the classic 802.11 approach). We also add the obtained result in [19, 21] for *SyncScan* and *SelectiveScan*, respectively. For comparison purposes, the traffic load for all nodes is divided into real-time and non real-time traffic with a ratio of 7.5/2.5. Other than errors caused by handoff occurrences, the real-time data packets are dropped also if the inter-frame delay exceeds 50ms. The simulation time for each traffic type is 10s (equivalent to about 2500 frames). Clearly, PSHP outperforms the other three schemes and the basic 802.11 as long as the traffic load is limited. The loss probability value of the new PSHP technique is divided by two compared to these obtained by *SyncScan* and *SelectiveScan* methods and by three of that accomplished by the standard 802.11 scheme.

In conclusion, periodic scanning also provides the means to make more intelligent choices when to initiate handoff. The new implementation can discover the presence of APs with stronger RSSIs even before the associated AP's signal has degraded below the threshold. In addition, the pre-scan phase does not affect the existing wireless traffic since the corresponding MS will carry out a pre-scan cycle after declaring the PSM mode to buffer related packets.

**5.2. Evaluation of the new add-on AP-selection heuristic**

As mentioned above we add new context-based parameters for the next AP choice when a handover is triggered in the network by a MS. The result technique is not dependent on the used handoff method. Thus, we integrate the new developed heuristic function with both the classic and the proposed switching algorithm. Specifically, in the standard 802.11 method the next AP selection will be performed after the scan phase on the found APs by choosing one based on the new objective function. Regarding the PSHP procedure this choice will be performed after each pre-scan cycle only on APs that belong the associated dynamic list. This function is also performed for both handoff Form 2 and Form 3. The only algorithm modification in PSHP Form 1 handoff process is that the objective function is performed only on listed APs that have an RRSI value greater than the actual RSSI measured between the MS and its actual AP. By adopting this condition we always maintain the main purpose of the PSHP which is an earlier selection of a new AP that offers a better link quality. Therefore, the modified PSHP will not choose automatically the first best AP in the list. However, it will select from existing AP that maximizes the objective function and also offers a better channel link quality. We set the same simulation parameters as given in Table 1. However, we add geographic constrains by influencing some MS-AP link qualities depending on AP initial positions and by introducing initial specific values for the CNX parameter to illustrate the already performed MS-journeys in the network and a random primary associations between MSs and the given set of APs. The simulated mobility model regarding the MS moves is no longer "Random Way Point". To be closer to realistic networks and to better assess our mechanism we switch to the "Random Direction" Mobility Model which forces mobile stations to travel to the edge of the simulation area before changing direction and speed. We choose this model because of its inclusion simplicity and instead of the "City Section" Mobility Model – which represents streets within a city. By including these constrain, we evaluated of the proposed heuristic combined with handoff schemes. In Figure 13 we resume the handoff occurrences for both classic and modified handoff schemes for the standard 802.11 and the PSHP techniques according to the traffic load. We set the simulation time to 10s for each considered traffic load. We point out a perceived reduction for handoff occurrences for both schemes when using the proposed heuristic procedure during the next AP selection. The produced results with the PSHP procedure are clearly enhanced in term of handoff count by integrating the new add-on heuristic technique. This is the use effect of the new objective function that accomplishes a better AP choice for the next inter-cell commutation, and consequently, improves the total number of handoff happening by reducing worse AP selections that was based only on RSSI-measurement decisions.

The detailed number of the various handoff forms related to the extended PSHP technique is shown in Figure 14. As well as in Figure 12, the vertical red and blue lines represent, respectively, the executed number of Form 1 handoffs and the count of handoffs taken under the second and third form (called also urgent handoffs). We figure out that handoffs Form 1 – performed when the RSSI value degrades below the $RSSI_{prev}$ threshold – are more triggered using the modified PSHP. We note that the proposed algorithm detects earlier the MS path and direction based on supplementary context-based information, and as a result, chooses quicker the best AP that improves the link quality and offers a continuous channel connection. Accordingly, 72% of accomplished handoffs are done under the first form of PSHP that decrease considerably the total latency and improves the link quality. As discussed before, data packets are dropped mainly by the handoff procedure and the violation of VoIP restrictions. Table 4 shows the data loss average probability for both classic 802.11 and PSHP approaches. As settled before the simulation time is 10s. The traffic load for MSs is equally combining real-time and non real-time traffic. The given results are the average of simulated values by varying the traffic load (from lower to higher loads).

Table 3. VoIP packet's loss Table 4.

| Scan Technique | Loss Probability |
|---|---|
| SyncScan [19] | 0.92 x1E-02 |
| Selective Scan [21] | 1.28 x1E-02 |
| APFH [23] | 0.72 x1E-02 |
| IEEE 802.11 handoff | 1.62 x1E-02 |
| New PSHP | 0.53 x1E-02 |

Table 4. Packet's loss with heuristic selection

| Scan Technique | Loss Probability |
|---|---|
| Standard 802.11 handoff | 1.62 x1E-02 |
| PSHP | 0.53 x1E-02 |
| IEEE802.11+heuristic selection | 0.78 x1E-02 |
| PSHP + heuristic selection | 0.32 x1E-02 |

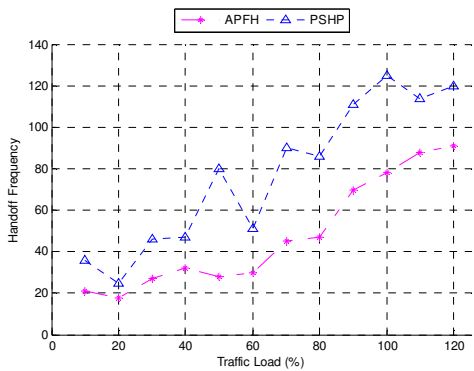

Figure 11. Handoff Frequency

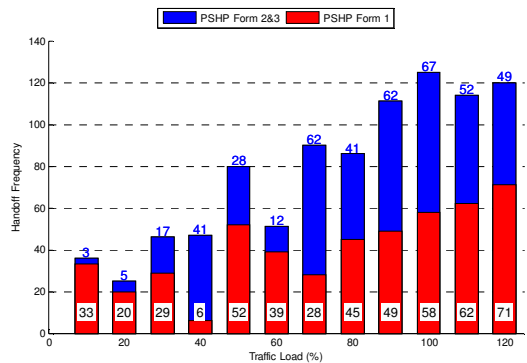

Figure 12. Occurrence of Handoff forms in PSHP

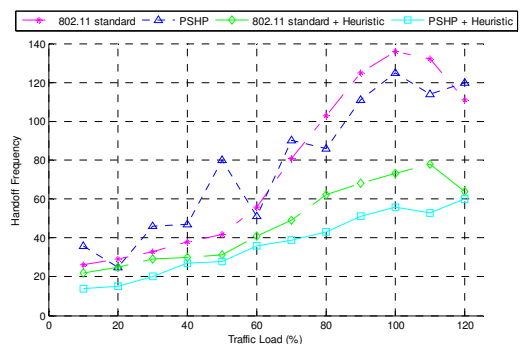

Figure 13. WLAN Handoff's frequency

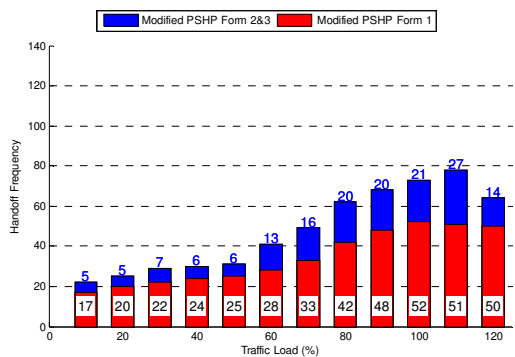

Figure 14. Handoff Occurrence in PSHP

We note that the modified PSHP version is outperforming the regular scheme. The reduced number of handoffs and also the high percentage of Form 1 handoffs lead to minimize the packet loss caused by handoff procedures. Thus, we can conclude that the loss probability value obtained by the new PSHP integrating the heuristic technique includes mainly dropped packets associated to a higher traffic load and not linked to the lack of respect of QoS constrains.

## 6. CONCLUSIONS

Mobile voice applications are currently the challenge for 802.11-based WLANs. One of the major impediments is the high cost of handoff as MSs room between APs in an infrastructure network. In this paper, we firstly presented a new technique, called PSHP, which reduces the delay and the traffic generated by the handoff process. As demonstrated, the continuous scanning PSHP technique offers significant advantages over other schemes by minimizing the time during which an MS remains out of contact with its AP and allowing handoffs to be made earlier and with more confidence. The result is a staggering 95% reduction of handoff latency compared to the typical procedure. As a second contribution we took into account additional network-based parameters to drive a better next-AP choice. This new add-on profit function is used to insert new factors reflecting resource availability, location, and other context-based information. Thus, the overall network performance is improved by electing from available APs, the one that increases the benefit of the next handoff occurrence.